\documentclass[final,10pt,times,3p]{elsarticle}
\usepackage{graphicx,epstopdf,amsmath,amssymb,braket,color,epsfig,epstopdf,geometry,amsthm}
\usepackage[colorlinks=true, citecolor=red, urlcolor=blue ]{hyperref}
\usepackage[table]{xcolor}
\usepackage{times,array,multirow}
\journal{Physics Letters A}

\begin{document}

\title{Deterministic Quantum Dense Coding Networks}

\author{Saptarshi Roy${}^1$, Titas Chanda${}^1$, Tamoghna Das${}^{1,2}$, Aditi Sen(De)${}^1$, and Ujjwal Sen${}^1$}


\address{${}^1$Harish-Chandra Research Institute, HBNI, Chhatnag Road, Jhunsi, Allahabad 211019, India \\
${}^2$Institute of Informatics, National Quantum Information Centre, Faculty of Mathematics, Physics and Informatics, \\
University of Gdan\'{s}k, 80-308 Gdan\'{s}k, Poland
}

\begin{abstract}
We consider the scenario of deterministic classical information transmission between
multiple senders and a single receiver, when they
a priori share a multipartite quantum state -- an attempt  towards building
a deterministic dense coding network.  Specifically, we prove that in the case
of two or three senders and a single receiver,   generalized
Greenberger-Horne-Zeilinger (gGHZ) states  are not beneficial for sending
classical information deterministically beyond the classical limit, except
when the shared state is the GHZ state itself. On the other hand,  three- and four-qubit 
generalized W (gW) states with specific  parameters as well as the four-qubit Dicke states
can provide  a quantum  advantage of sending the information in deterministic
dense coding. Interestingly however,
numerical simulations in the three-qubit scenario reveal that the 
percentage of states from the GHZ-class that are  deterministic dense
codeable is higher than that of  states from the W-class.
\end{abstract}


\maketitle





\section{Introduction}
The rapid development of quantum information science is largely due to  discoveries of communication protocols \cite{BB84,  cryptoEkert, Bennett_DC, teleportation, repeater, reviewcomm} by using entangled quantum states \cite{HHHHreview}. 
Their successful realizations in physical systems like photons \cite{photonexp}, ions \cite{ionexp}, superconducting qubits \cite{supercondexp}, nuclear magnetic resonance (NMR) \cite{NMR} etc. also make the field attractive. When a priori an entangled state is  shared between sender(s) and receiver(s), tasks of 
communication protocols can broadly be classified in two categories  -- classical information 
\cite{Bennett_DC, BB84, cryptoEkert}  and quantum state transfer \cite{teleportation}.
 The former, without the security issue during the  transmission of information, is known as the quantum dense coding protocol (DC)   \cite{Bennett_DC} which is the main theme of this rapid communication. 
The quantum DC protocol  has been experimentally implemented with photons \cite{DC_exp} and later with NMR \cite{NMR_DCexp}, trapped ions \cite{ion_DCexp}, and also in continuous variable systems \cite{DC_expCV}.

The original DC protocol describes the advantage to send the information of $N$ possible outcomes of a classical random variable, say $X$, when encoded in a quantum state from a single sender (Alice) to a single receiver (Bob). Bennett and Wiesner \cite{Bennett_DC} have shown that if Alice and Bob share the maximally entangled singlet state, $|\psi^-_{AB}\rangle = \frac{1}{\sqrt{2}} (|01\rangle - |10\rangle)$, Alice can transform the state into four possible orthogonal states by acting local unitaries on her part and
can send $\log_2 4 = 2$ bits of classical information by sending only a single spin-$1/2$ particle, i.e., a two-dimensional system.
If the initially shared state is $|\phi_{AB}\rangle =a |01\rangle + b |10\rangle$, where $a, b \in \mathbb{C}$ with $\mathbb{C}$ being the set of complex numbers  and $|a|^2 + |b|^2 = 1$ with $a \neq b$ (both non-zero), or an arbitrary state, $\rho_{AB}$,  Alice can no longer create orthogonal output states by performing unitaries and hence the receiver gets less information. 
In the asymptotic limit, when many copies of  $\rho_{AB}$ are provided, the amount of maximal classical information transferred on an average is the dense coding capacity ($C$) \cite{capacity_single, capacity_multi},  given by 
$C(\rho_{AB}) = \log_2 d_A + \mbox{max}\{S(\rho_B) - S(\rho_{AB}), 0\},$
where $d_A$ is the dimension of the Hilbert space of the sender's subsystem, $S(\sigma) = -\text{tr}(\sigma \log_2 \sigma)$ is the von Neumann entropy of $\sigma$, and $\rho_{B} = \text{tr}_A(\rho_{AB})$ is the reduced density matrix of the receiver's subsystem.  The first term is the classical limit for information transfer, while the remaining terms quantify the quantum advantage in DC. Clearly, in the case of  pure states, the entanglement content of the shared state  \cite{bennett_ent}  and the quantum advantage of DC capacity is equal. 

Instead of considering an asymptotic way of transferring classical bits which also is probabilistic in nature,  
we deal with  a DC scheme in a single-copy level, using a shared non-maximally entangled pure state, where Alice encodes the information by performing unitary operations on her part in such a way that upon receiving the entire system, Bob can  always distinguish the output states  without any error, i.e., deterministically, by  performing global measurements. Such protocol for a single sender and  a single receiver  was introduced in   Ref. \cite{DDC_Benny}, and referred to as the
 deterministic  dense coding (DDC) protocol
\cite{DDC_others}. 
  Since the protocol is at the single-copy level, it is also important from an experimental point of view \cite{DDC_exp_propose}. 
In DDC, Alice's aim is to find unitary operators, $\{U_i^A\}$, such that   mutually orthogonal states  can be created by applying $\{U_i^A\}$ on her part of $|\psi_{AB}\rangle$, thereby distinguishing them by Bob using global measurements.   
%
If $|\psi_{AB}\rangle \in \mathbb{C}^d\otimes \mathbb{C}^d$, 
where $d$ is the dimension of each subsystem, the classical limit of the alphabet-size of the message is $d$, while $\ket{\psi_{AB}}$ is said to be deterministically dense codeable  if the maximal number of orthogonal unitary operators, $N^{\psi}_{max}$, is greater than  $d$. 
It was proven that the entire family of pure states in $\mathbb{C}^2\otimes \mathbb{C}^2$  except the maximally entangled state  \cite{DDC_Benny} is useless for deterministic dense coding.
Till now, all the studies on DDC are restricted to a single sender and a single receiver (cf. \cite{DDC_multiparty}), although the importance of  building a communication  protocol between several senders and several receivers is unquestionable. In this work, we address the question of building a DDC network between several senders and a single receiver. 
Interestingly, we show that DDC is possible even with two-level systems already if one increases the number of senders to two.
We first prove that the DDC protocol with quantum advantage is \emph{not} possible when the shared state is a 
generalized GHZ state with two or more than two senders and a single receiver except when it is a GHZ state for which DDC and DC attain the maximum capacities. 
We show that the DDC scheme can be executed by using the generalized W states beyond the classical limit. We also perform a comparison between the states from the GHZ- and the W-classes according to their usefulness in  DDC. 
Moreover, we comment that the maximal number of unitaries cannot reach $d^{M+1} - 1$, when a \((M +1)\)-party  state is shared between \(M\) senders and a single receiver, each having dimension \(d\) (cf. \cite{Duan_d2} for two-qudit states). 
%



\section{Deterministic Dense Coding Network: Many Senders and A Single Reciever}
We now extend the deterministic dense coding protocol to multiple senders and a single receiver, situated in distant locations.
Let us consider a $(M+1)$-party pure state $\ket{\psi_{S_1 S_2...S_M R}}$  shared between the $M$ senders, $S_1, S_2, ..., S_M$, and a single receiver, $R$. A set of arbitrary local unitary operators, $\{U_i^{S_k}\}$ is performed by each sender, $S_k$.
Our task is to find out the maximal number of unitaries of the form $\{\bigotimes_k U_i^{S_k}\}$ such that 
the members of the
set of output states
 $\{ \bigotimes_k U_i^{S_k} \otimes \mathbb{I}^{R} \ket{\psi_{S_1 S_2...S_M R}} \}$,  sent to the receiver, are
 mutually orthogonal to each other. Hence, we find $\{U^{S_k}_i\}$, satisfying
\begin{eqnarray}
\bra{\psi_{S_1 S_2...S_M R}} \left(\bigotimes_k {U_i^{S_k {\dagger}}} \otimes \mathbb{I}^{R}\right)
\left(\bigotimes_{k'} {U_j^{S_{k'}}} \otimes \mathbb{I}^{R}\right) \ket{\psi_{S_1 S_2...S_M R}}  = \delta_{ij},
\label{Eq:DDC_main1}
\end{eqnarray}
or, alternatively
\begin{eqnarray}
\text{tr}\left(\big(\bigotimes_k {U_i^{S_k {\dagger}}} \big)\rho_{S_1 S_2 ...S_M} \big(\bigotimes_{k'} {U_j^{S_{k'}}} \big) \right) = \delta_{ij},
\label{Eq:DDC_main2}
\end{eqnarray}
where $\rho_{S_1 S_2 ...S_M} = \text{tr}_{R} \big(\ket{\psi_{S_1 S_2 ... S_M R}} \bra{\psi_{S_1 S_2 ... S_M R}}\big)$ is the reduced density matrix of all the senders' subsystems for a given state $\ket{\psi_{S_1 S_2 ... S_M R}}$. The aim is to find  
the maximal number of such unitaries, $N^{\psi}_{max}$, which will define the alphabet-size of the message that senders can send.
We can always choose the identity operator $\bigotimes_k \mathbb{I}_{S_k}$ on the Hilbert space of the senders as one of the members of the above set of orthogonal unitary matrices $\{\bigotimes_k U_i^{S_k}\}$.
The task then reduces to find remaining $N^{\psi}_{max} - 1$ number of unitary matrices, satisfying Eq. (\ref{Eq:DDC_main2}), either analytically or by numerical simulations. It is noteworthy to mention that, in general, $N^{\psi}_{max}$ lies in the range $[d^{M}, d^{M+1}]$, where $d^{M}$ is the classical limit and $d^{M+1}$ is the quantum limit of the alphabet-size. For a given state, $|\psi\rangle$, if we find that $N^{\psi}_{max} > d^M$, we conclude that the state has quantum advantage in DDC.

Let us restrict ourselves to two senders and a single receiver. They now share a three-qubit pure state $\ket{\psi_{S_1 S_2 R}}$ and each sender performs a two-dimensional unitary operator given by
\begin{eqnarray}\label{Eq:unitary}
U_i^{S_k}=
  \left( {\begin{array}{cc}
   \cos\theta_i^{S_k}e^{i x_{i}^{S_k}} &  - \sin\theta_i^{S_k}e^{i y_{i}^{S_k}}\\
   \sin\theta_i^{S_k}e^{- i y_{i}^{S_k}} & \cos\theta_i^{S_k}e^{-i x_{i}^{S_k}} \\
  \end{array} } \right),
  \label{eq:unitary}
\end{eqnarray}
where $\theta_i^{S_k} \in [0, \pi]$ and $x_i^{S_k}, y_i^{S_k} \in [0, 2\pi]$.
Notice that we have chosen $U_i^{S_k}$ as  an element of \emph{{SU}}$(2)$, since 
any arbitrary value of the determinant does not contribute to the orthogonality condition except a global phase. 
In this case, Eq. (\ref{Eq:DDC_main2}) reduces to
 \begin{equation}
 \text{tr}\big(({U_{i}^{S_1 \dagger} \otimes U_{i}^{S_2 \dagger}})\rho_{S_1 S_2} ({U_j^{S_1} \otimes U_j^{S_2}}) \big) = \delta_{ij}.
\label{Eq:DDC_main3}
\end{equation}  
We will show that unlike two-qubit states, for three-qubit pure states, the solution of Eq. (\ref{Eq:DDC_main3}) exists, thereby ensuring quantum advantage by DDC scheme. A similar observation can also be made for a higher number of senders.

\section{DDC: GHZ-class vs. W-class}
Let us first consider two important families of three-qubit states. They are the generalized GHZ (gGHZ) states \cite{GHZ}, given by 
\begin{eqnarray}
|gGHZ_{S_1 S_2 R}\rangle = \sqrt{\alpha} |00 0\rangle + \sqrt{1- \alpha}~e^{i \mu}|11 1\rangle,
\label{eq:gGHZ1}
\end{eqnarray}
 where $\alpha \in [0,1]$ and $\mu \in [0,2\pi)$, and the generalized W (gW) states \cite{Wclass},
\small
\begin{eqnarray} 
\ket{gW_{S_1 S_2 R}} = \sqrt{\alpha}\ket{001} + \sqrt{\beta} \ket{010} + \sqrt{1 - \alpha -\beta} \ket{100}
\label{eq:gW1}
\end{eqnarray}\normalsize
 with $\alpha, \beta \in [0, 1]$, and $ \alpha + \beta \leq 1$. For $\alpha = \frac{1}{2}$ in Eq. (\ref{eq:gGHZ1}), we get the well known GHZ state, while in Eq. (\ref{eq:gW1}) we have the W state for $\alpha = \beta = \frac{1}{3}$. The gGHZ states and the gW states are well known subsets (of measure zero) of two SLOCC (stochastic local operations and classical communication) inequivalent classes of three-qubit pure states \cite{dur_vidal_cirac}, namely, the GHZ-class \cite{GHZ_class_state} and the W-class \citep{W_class_state} respectively. The set of tripartite states, that can be converted into the GHZ state using only SLOCC, defines the GHZ-class, whereas the W-class contains all the tripartite states that can be converted, by means of SLOCC, into the W state. These two classes are inequivalent in the sense that one cannot convert, with finite probability, a member of the GHZ-class into a member of the W-class, or vice-versa, using SLOCC.
We will prove that although the gGHZ states (subset of GHZ-class) is not good for DDC, the quantum advantage of DDC is possible using the gW states (subset of W-class) by showing $N_{max}$ beyond the classical limit.

\subsection{No DDC for generalized Greenberger-Horne-Zeilinger states}
Suppose the shared state is unentangled in sender and receiver bipartition, then the maximum amount of information that the two senders, each having two-dimensional systems can send to the receiver is two bits. 
Moreover,  the capacity of dense coding with the GHZ state reaches its maximum value, implying successful implementation of DDC protocol with $N_{max}^{GHZ} = 8$ \citep{capacity_multi}. Let us consider DDC by using gGHZ states with $\alpha \neq \frac{1}{2}$.
For several reasons, including that in Theorem 1 below, the GHZ state is considered to be the ``maximally entangled'' among gGHZ states. 
See Ref. \cite{GHZ_distill} in these regards. We therefore refer to gGHZ states with $\alpha \neq \frac{1}{2}$ as ``non-maximally entangled'' gGHZ states.  

\noindent{\textbf{Theorem 1.}} \emph{ Non-maximally entangled generalized Greenberger-Horne-Zeilinger states are not useful for deterministic dense coding with two senders and a single receiver.} 

\noindent {\bf Proof.} 
The most general local encoding at the sender's end is the set of unitaries 
$\{U^{S_1}_i \otimes U^{S_2}_i \otimes \mathbb{I}_2^{R} \}$,
where $\mathbb{I}_2^{R}$ is a $2\times2$ identity matrix and $U_i^{S_k}$'s are given by Eq. (\ref{Eq:unitary}) which includes the identity.

To show that a gGHZ state is capable for DDC beyond the classical limit, we have to find more than four unitaries which include $\mathbb{I}^{S_1}_{2} \otimes \mathbb{I}^{S_2}_{R}$. We now look for unitary operators, which are orthogonal to the identity as well as orthogonal among themselves.

Given a general unitary $U^{S_1}\otimes U^{S_2} \otimes \mathbb{I}_2^R$, Eq. \eqref{Eq:DDC_main1} demands
$\Bra{gGHZ}  U^{S_1}\otimes U^{S_2} \otimes \mathbb{I}_2^R \Ket{gGHZ} = 0,$
i.e., $\big(\alpha e^{i(x^{S_1} +x^{S_2})} +(1-\alpha)e^{-i(x^{S_1} +x^{S_2})}\big)\cos\theta^{S_1}\cos\theta^{S_2} = 0$.
For $\alpha \neq \frac{1}{2}$,  we have
$\alpha e^{i(x^{S_1} +x^{S_2})}  + (1-\alpha)e^{-i(x^{S_1} +x^{S_2})} \neq  0,$
and hence
$\cos\theta^{S_1}\cos\theta^{S_2} = 0$. 
The solution of the equation gives rise to the following three classes of unitaries:
\begin{center}
\begin{tabular}{ |c|c|c| }
\hline
$Class$ & $\theta^{S_1}$ & $\theta^{S_2}$ \\ 
\hline
$\mathcal{C}_1$ & $\pi /2$ & arbitrary  \\ 
  
$\mathcal{C}_2$ & arbitrary & $\pi /2$ \\
  
$\mathcal{C}_3$ & $\pi /2$ & $\pi /2$ \\
 \hline
\end{tabular}
\end{center}

\noindent It is clear that classes $\mathcal{C}_1$ and $\mathcal{C}_2$ are equivalent and all the proofs given below for class $\mathcal{C}_1$ also holds for $\mathcal{C}_2$. 
We now list the possible cases which occur when unitaries are chosen from $\mathcal{C}_1$ or $\mathcal{C}_2$ or $\mathcal{C}_3$: 

{\bf Case 1.} We cannot make two or more unitaries from class $\mathcal{C}_3$ orthogonal.
Consider two unitaries $(U^{S_1}_{a_3} \otimes U^{S_2}_{a_3} \otimes \mathbb{I}_2^R)$ and $(U^{S_1}_{b_3} \otimes U^{S_2}_{b_3} \otimes \mathbb{I}_2^R)$ from class $\mathcal{C}_3$. The orthogonality condition reads as
$\alpha e^{i(y^{S_1}_{a_3} + y^{S_2}_{a_3} + y^{S_1}_{b_3} + y^{S_2}_{b_3})}  + (1-\alpha)e^{-i(y^{S_1}_{a_3}  + y^{S_2}_{a_3} + y^{S_1}_{b_3} + y^{S_2}_{b_3})} = 0$, 
which is \textit{not possible} when $\alpha \neq \frac{1}{2}$. 

{\bf Case 2.} We cannot make more than two unitaries from class $\mathcal{C}_1$ (or $\mathcal{C}_2$) orthogonal.
Let us consider two unitaries $(U^{S_1}_{a_1} \otimes U^{S_2}_{a_1} \otimes \mathbb{I}_2^R)$ and $(U^{S_1}_{b_1} \otimes U^{S_2}_{b_1} \otimes \mathbb{I}_2^R)$ from class $\mathcal{C}_1$ and their arbitrary angle variables are $\theta^{S_2}_{a_1}$ and $\theta^{S_2}_{b_1}$ respectively. From the orthogonality condition $\Bra{gGHZ} (U^{S_1}_{a_1} \otimes U^{S_2}_{a_1} \otimes \mathbb{I}_2^R)^\dagger (U^{S_1}_{b_1} \otimes U^{S_2}_{b_1} \otimes \mathbb{I}_2^R) \Ket{gGHZ} = 0$, we get
\small\begin{eqnarray}
\big( \alpha e^{i(x^{S_1}_{a_1} + y^{S_2}_{a_1} + x^{S_1}_{b_1}  + y^{S_2}_{b_1})}  
+ (1-\alpha)e^{-i(x^{S_1}_{a_1} + y^{S_2}_{a_1} + x^{S_1}_{b_1} + y^{S_2}_{b_1})}\big) \cos\theta^{S_2}_{a_1}\cos\theta^{S_2}_{b_1} 
+ \ \big( \alpha e^{i(y^{S_1}_{a_1} + y^{S_2}_{a_1} + y^{S_1}_{b_1} + y^{S_2}_{b_1})}  
+ (1-\alpha)e^{-i(y^{S_1}_{a_1} + y^{S_2}_{a_1} + y^{S_1}_{b_1} + y^{S_2}_{b_1}}) \big)`
 \sin\theta^{S_2}_{a_1}\sin\theta^{S_2}_{b_1} = 0.
\label{eq:eq3}
\end{eqnarray} \normalsize
Now Eq. (\ref{eq:eq3}) is satisfied when
$(x^{S_1}_{a_1} + x^{S_1}_{b_1}) = (y^{S_1}_{a_1} + y^{S_1}_{b_1})$ and
$\cos(\theta^{S_2}_{a_1})\cos(\theta^{S_2}_{b_1}) + \sin(\theta^{S_2}_{a_1})\sin(\theta^{S_2}_{b_1}) = 
\cos(\theta^{S_2}_{a_1}-\theta^{S_2}_{b_1}) = 0$
hold simultaneously.
We can easily adjust the phases of the unitaries to make the first condition valid and choose
\begin{equation}
\theta^{S_2}_{b_1} = \theta^{S_2}_{a_1} \pm \frac{\pi}{2},
\label{eq:eq4}
\end{equation}  
so that Eq. (\ref{eq:eq3}) holds. Therefore, we can choose two orthogonal unitaries from class $\mathcal{C}_1$. It is to be noted that none of the above angle variables can be equal to zero, as it would require the other to be $\pi/2$, which does not happen for the unitaries of class $\mathcal{C}_1$.
Now suppose we can choose another unitary from class  $\mathcal{C}_1$, $(U^{S_1}_{c_1} \otimes U^{S_2}_{c_1} \otimes \mathbb{I}_2)$ with the arbitrary angle  $\theta^{S_2}_c$. Similar to the condition given in  Eq. (\ref{eq:eq4}), to make this third unitary simultaneously orthogonal to the first two, we require   
\begin{eqnarray}\label{eq5}
\theta^{S_2}_{c_1} = \theta^{S_2}_{a_1} \pm \dfrac{\pi}{2}, \quad \textrm{and} \quad \theta^{S_2}_{c_1} = \theta^{S_2}_{b_1} \pm \dfrac{\pi}{2}. 
\end{eqnarray}
However such a choice is a contradiction to Eq. (\ref{eq:eq4}). Hence 3 orthogonal unitaries from class $\mathcal{C}_1$ cannot be chosen. A similar proof also holds for class $\mathcal{C}_2$. 

{\bf Case 3.} If we make two unitaries from class $\mathcal{C}_1$ orthogonal, then they cannot be orthogonal to any unitary from class $\mathcal{C}_3$.
For a unitary $(U^{S_1}_{a_1} \otimes U^{S_2}_{a_1} \otimes \mathbb{I}_2^R)$ from class $\mathcal{C}_1$ with arbitrary angle variable $\theta^{S_2}_{a_1}$ to be orthogonal to a unitary $(U^{S_1}_{b_3} \otimes U^{S_2}_{b_3} \otimes \mathbb{I}_2^R)$ from class $\mathcal{C}_3$, one requires that
\small
\begin{eqnarray}
\big(\alpha e^{i(y^{S_1}_{a_1} + y^{S_2}_{a_1} + y^{S_1}_{b_3} + y^{S_2}_{b_3})}  + (1-\alpha)e^{-i(y^{S_1}_{a_1} + y^{S_2}_{a_1} + y^{S_1}_{b_3} + y^{S_2}_{b_3})}\big)   \sin(\theta^{S_2}_{a_1})= 0.
\label{eq:eq6}
\end{eqnarray}
\normalsize
For $\alpha \neq \frac{1}{2}$, the above condition is true only when, $\sin\theta^{S_2}_{a_1} = 0 \text{ which implies } \theta^{S_2}_{a_1} = n\pi,\text{ where } n=0,1,2,3 ...$ But we have shown that if class $\mathcal{C}_1$ permits two orthogonal unitaries, their angle variables cannot be equal to zero (Eq. (\ref{eq:eq4})). 

{\bf Case 4.} We now show that it is possible to make two unitaries from class $\mathcal{C}_1$, one from class $\mathcal{C}_2$ orthogonal and vice versa, but not more than that.
As before, a unitary $(U^{S_1}_{a_1} \otimes U^{S_2}_{a_1} \otimes \mathbb{I}_2^R)$ from class $\mathcal{C}_1$ with arbitrary angle variable $\theta^{S_2}_{a_1}$ is orthogonal to a unitary $(U^{S_1}_{b_2} \otimes U^{S_2}_{b_2} \otimes \mathbb{I}_2^R)$ from class $\mathcal{C}_2$ with arbitrary angle variable $\theta^{S_1}_{a_2}$, implies that
\begin{eqnarray}
\big(\alpha e^{i(y^{S_1}_{a_1} + y^{S_2}_{a_1} + y^{S_1}_{b_2} + y^{S_2}_{b_2})}  + (1-\alpha)e^{-i(y^{S_1}_{a_1} + y^{S_2}_{a_1} + y^{S_1}_{b_2} + y^{S_2}_{b_2})}\big)  \sin\theta^{S_2}_{a_1}\sin\theta^{S_1}_{b_2}= 0.
\end{eqnarray}
Again for $\alpha \neq \frac{1}{2}$, we get
\begin{equation}
\sin\theta^{S_2}_{a_1}\sin\theta^{S_1}_{b_2}= 0 
\label{eq:eq7}
\end{equation}
From Eq. \eqref{eq:eq4} it follows that if we choose two orthogonal unitaries from class $\mathcal{C}_1$, their angle variables must be non-zero. Hence the only way Eq. (\ref{eq:eq7}) can hold, is when $\theta^{S_1}_{b_2} = 0$, which implies a single unitary from class $\mathcal{C}_2$. It was proven in Case 3 that if we make two unitaries from the class $\mathcal{C}_1$ orthogonal, they will not be orthogonal to any unitary from the class $\mathcal{C}_3$. Therefore, in this case also the maximal number of orthogonal unitaries apart from the identity is three.
The situation of two unitaries from class $\mathcal{C}_2$, one from class $\mathcal{C}_1$ is exactly the same.

{\bf Case 5.} At most a single unitary from each class can be made orthogonal to each other.
Let us choose unitaries $(U^{S_1}_{\nu} \otimes U^{S_2}_{\nu} \otimes \mathbb{I}_2^R)$, $\nu = a_1,b_2$ with $\theta^{S_1}_{a_1}$ and $\theta^{S_2}_{b_2}$ from classes $\mathcal{C}_1$
and $\mathcal{C}_2$ respectively. And similarly, $(U^{S_1}_{c_3} \otimes U^{S_2}_{c_3} \otimes \mathbb{I}_2)$ from $\mathcal{C}_3$. Mutual orthogonalities of these three unitaries imply $\theta_{a_1}^{S_1} = \theta_{b_2}^{S_2} = 0$, thereby satisfying both Eqs. \eqref{eq:eq6} and \eqref{eq:eq7}. Previous cases ensure that it is not possible to choose any more orthogonal unitaries from any of the classes. Therefore, the maximal number of orthogonal unitaries apart from the identity is also restricted to three.



Exhausting all the cases, we conclude that for the gGHZ states with $\alpha \neq \frac{1}{2}$, $N_{max}(|gGHZ_{S_1S_2R}\rangle) = 4$ if one includes identity and hence the proof. \hfill $\blacksquare$

It is interesting to note that the GHZ state having maximum genuine multiparty entanglement, has maximum DC and DDC capacities. The other states in this family, with an infinitesimally lower genuine multipartite entanglement are completely useless for DDC implying a discontinuity in their DDC capacity.

\subsection{DDC for generalized W states}
Let us now show that when the shared state is $|gW_{S_1S_2R}\rangle$, $N_{max}^{gW} > 4$ for some parameter values of $\alpha \text{ and } \beta$. For a given shared state $\ket{\psi_{S_1 S_2 ... S_M R}}$, we have to find $N_{max}^{\psi} -1$ orthogonal unitaries excluding the identity operator, each of which has $M(d^2 -1)$ unknown parameters, where $d$ is the dimension of each sender, and $M$ is the number of senders. In a three-qubit scenario, with two senders and a single receiver, the number of unknown parameters for each unitary is $ 6 $.
On the other hand, the orthogonality condition between different unitaries, as given in Eq. (\ref{Eq:DDC_main3}), gives us $\frac{1}{2}N_{max}^{\psi}(N_{max}^{\psi} - 1)$ equations. Therefore, we have a multivariate root finding problem of  $\frac{1}{2}N_{max}^{\psi}(N_{max}^{\psi} - 1)$ 
equations with $M(N_{max}^{\psi}-1)(d^2 - 1)$ unknowns, which can be solved by standard root-finding algorithms \cite{multi_root}. If there exists no solution for a particular shared state and a given $N_{max}^{\psi}$,  we then conclude that for that shared state, we can not have $N_{max}^{\psi}$ number of unitaries, including identity, that satisfy Eq. (\ref{Eq:DDC_main3}). However, if for a shared state, we can have $N_{max}^{\psi}$ number of orthogonal unitaries, we must have a solution for the problem, and the algorithm must converge for at least one starting point. In our simulations, we observe that either the algorithm fails to converge or it converges for every single random starting point, for fixed $N_{max}^{\psi}$ and a particular shared quantum state. We say that the algorithm converges, if the LHS of Eq. (\ref{Eq:DDC_main3}) is less than $10^{-5}$.

\begin{figure}
\begin{center}
\includegraphics[width=0.6\linewidth]{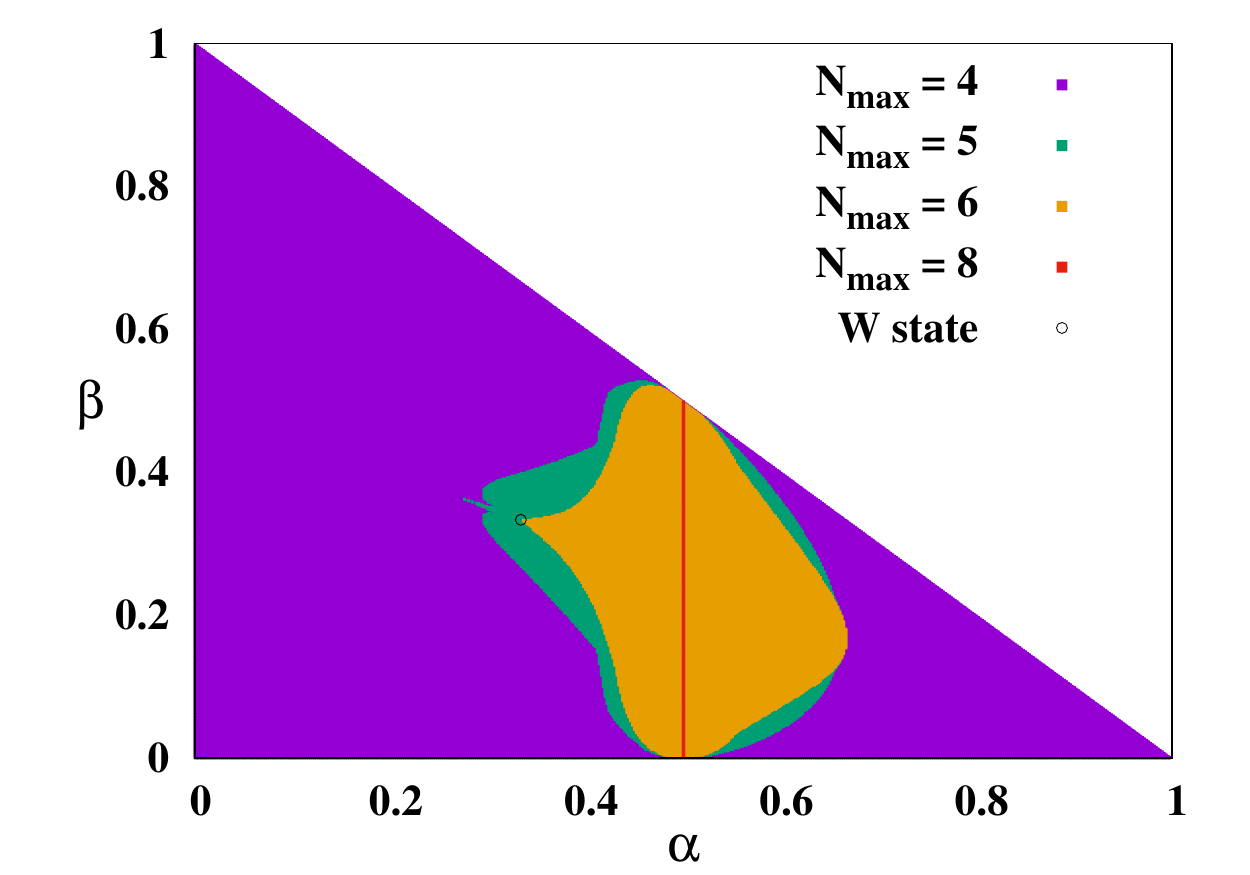}
\caption{Map of maximal number of unitaries $N_{max}^{gW}$ for gW states with respect to their parameters $\alpha \text{ and } \beta$. All quantities are dimensionless.}
\label{fig:gW}
\end{center}
\end{figure}

\begin{figure*}
\includegraphics[width=1.05\linewidth]{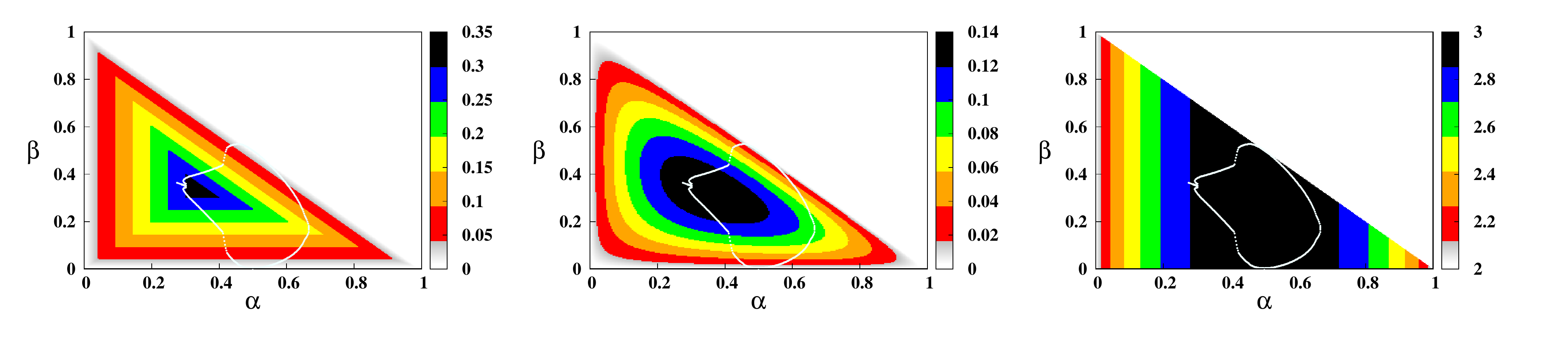}
\caption{Maps of generalized geometric measure as genuine multipartite
entanglement measure \cite{ent_multi}, squared negativity monogamy score \cite{neg_mono}, and dense coding
capacity \cite{capa_def} for generalized W states. The horizontal and vertical axes are the
same as in Fig. \ref{fig:gW}. The first two quantities are in ebits while
the last one is in bits. The white line indicates the boundary of states, inside of which the states are good for deterministic dense coding.  }
\label{fig:gW_qc}
\end{figure*}

It was known that there exists a  $\ket{gW_{S_1 S_2 R}}$ state which shows perfect dense coding for $\alpha = \frac{1}{2} \text{ and }\forall \beta $  \cite{pati_pankaj} with $N^{gW}_{max} = 8$ as seen in Fig. \ref{fig:gW}. 
By numerical simulation, we find that for certain values of $\alpha \text{ and } \beta$, $N_{max}^{gW} > 4$. Specifically, when $0.276 \leq \alpha \leq 0.362$, for some specific values or ranges of $\beta$, we show that DDC is possible with quantum advantage by sharing the corresponding three-qubit gW state.
In Fig. \ref{fig:gW}, we map numerically-obtained values of $N^{gW}_{max}$ with the parameters  $\alpha$ and $\beta$ which clearly depicts the quantum advantage of DDC. 
For the well-known W state, i.e., $\alpha = \beta = \frac{1}{3}$, we get $N_{max}^{W} = 6$. Surprisingly, we were unable to find any states from the set of gW states, that can have $N_{max}^{gW} = 7$. 

To address any possible connection between DDC capacity ($N_{max}$) with multiparty entanglement, we study several physical quantities, multipartite in nature.  They are  computable genuine multipartite
entanglement measure, known as generalized geometric measure  \cite{ent_multi},
monogamy of squared negativity  \cite{neg_mono},  and dense coding
capacity (which is a generalization of dense coding capacity mentioned earlier) \cite{capa_def} for the generalized W state
as depicted in Fig. \ref{fig:gW_qc}. The states lying inside the white line
in  figures  are all deterministic dense codeable, having quantum
advantage. It is clear that all these quantities give sufficient condition
for DDC. 

Moreover, looking at Fig. \ref{fig:gW}, one would be tempted to conclude that 
the decrease in number of orthogonal unitaries can be understood by the continuity argument. However, this is not  correct because such argument then implies that the gGHZ states, close to the GHZ state in the parameter space ($\alpha \sim 1/2$) are also capable for  deterministic dense coding which is not true. Note that, such continuity argument is in general valid for entanglement content as well as for the capacities of probabilistic dense coding in the sender:receiver bipartition for multiparty pure states. 
Therefore, the
general multipartite properties which are responsible for deterministic
dense coding beyond the classical limit still remains an open question.

\subsection{GHZ-class vs. W-class}
In our treatment, we generate $2.5 \times 10^{6}$ random pure states Haar uniformly from the GHZ-class \cite{dur_vidal_cirac, GHZ_class_state} as well as the W-class \cite{dur_vidal_cirac, W_class_state}, and obtain maximal number of orthogonal unitaries, $N_{max}$, for each of those states. Interestingly, unlike gGHZ and gW states, we observe that states from the GHZ-class have more chances to possess quantum advantages in DDC than the states from the W-class. Specifically, for GHZ-class, approximately $18.02 \%$ states have $N_{max}$ more than the classical limit $4$, while it is only about $2.65 \%$ of states from the W-class for which $N_{max} > 4$. We find that in the case of states from the GHZ-class, the number of states having $N_{max} = 6$ is higher than the states with $N_{max} =5$. The scenario is exactly the opposite for the W-class states. 
Another interesting point to note here is that if one considers a one parameter family of the W-class, given by
$\ket{\psi_{W_s}} = \sqrt{1-a} \ket{W} + \sqrt{a}\ket{000}$,
where $\ket{W}$ is the well known W state mentioned earlier, we observe that $N_{max}(\ket{\psi_{W_s}}) = 5$ for $0 < a < 0.035$, thereby indicating a very small set of states from the family that is deterministic dense codeable. It possibly suggests that among the W-class states, most of the states which are deterministic dense codeable actually belong to the generalized W states.
A detailed percentage distribution of states having non-classical DDC capacity is given in Table \ref{table_1}. 


To summerize, the set of gGHZ is not equivalent to the GHZ-class, but the former is a known subset (of measure zero) of the later. A similar relation holds between the set of gW states and W-class states. We find that gGHZ states are not good for DDC, while gW states can have quantum advantage in DDC. However, overall performance of GHZ-class is better than the W-class in terms of the percentage of states with DDC capability beyond the classical limit. Although it apparently looks counter-intuitive, there is nothing contradictory as one cannot expect measure zero subsets to mimic the features of the entire set. But the complete role reversal is definitely surprising and interesting.

\begin{table}
\begin{center}
\begin{tabular}{||c|c|c||}
\hline
SLOCC  & \multicolumn{2}{|c|}{ $N_{max} $} \\
\cline{2-3}
classes & 5 & 6 \\ 
\hline
GHZ-class & $8.25 \%$ & $9.77 \%$ \\ 
\hline
W-class & $1.57 \%$ & $1.08 \%$ \\
\hline
\end{tabular}
\end{center}
\caption{Percentage of states from the GHZ- and W-classes with a given $N_{max}$.}
\label{table_1}
\end{table}

It is important to note that we have not found a single state that accepts $N_{max} = 7$ for DDC with 2 senders and a single receiver.
It was shown that in case of a single sender and a single receiver, the maximal number of unitary cannot take the value $d^2 -1$ with $d$ being the dimension of the sender's subsystem. Note however, that if a shared bipartite state is in $d_S \otimes d_R$, $d_S$ and $d_R$ being the dimensions of the sender's and receiver's subsystem respectively, with $d_S > d_R$, we find that there does not exist any such no-go theorem, and $N_{max}$ can have all the values in the range $[d_S, d_S d_R]$ \cite{notun_kaj}.

In a multipartite scenario, after performing extensive numerical simulations with higher dimensional systems as well as higher number of senders, we are tempted to make the following conjecture (cf. \cite{DDC_Bourdon}):

\noindent \textbf{Conjecture 1.} \emph{If an $(M+1)$-party quantum state is shared between $M$ senders and a single receiver, where each party has dimension $d$,  the maximal number of unitaries $N_{max}$, given by Eq. (\ref{Eq:DDC_main2}), cannot be equal to $d^{M+1} - 1$.} \\
This conjecture is clearly in agreement with our findings that $N_{max} \neq 7$ for two senders each having two dimensions.
If  an $(M+1)$-party quantum state is shared between $M$ senders and a single receiver, where each party has dimension $d$, we can find at most $d^{M}$ orthogonal states that can be distinguished by the receiver, provided there is no entanglement between senders and the receiver, while for the perfect DC, we can find $d^{M+1}$ orthogonal states. The conjecture and numerical findings indicate that for DDC with quantum advantage, we can have $N_{max}$ in $[d^{M}, d^{M+1}]$ except $N_{max} = d^{M+1} - 1$.

\section{DDC with higher number of senders}
Let us move to the situation of a four-qubit quantum state, where the 
number of senders is now increased to three.
For such a scenario, one can ask whether the DDC using the gGHZ state, $
|gGHZ_{S_1 S_2 S_3 R}\rangle = \sqrt{\alpha} |0000\rangle + \sqrt{1- \alpha}~e^{i \mu}|1111\rangle,
$
 where $\alpha \in [0,1]$ and $\mu \in [0,2\pi)$, shared between three senders, $S_1$, $S_2$, and $S_3$ and a single receiver, $R$,
is possible beyond the classical limit. In this case also, we arrive at the following theorem:

\noindent{\textbf{Theorem 2.}} \emph{Non-maximally entangled generalized Greenberger-Horne-Zeilinger states are not useful for deterministic quantum dense coding with three senders and a single receiver.}

\noindent{\bf Proof:} 
The proof is in a similar spirit to Theorem 1. In this case, we have to show $N_{max}^{gGHZ}$ cannot go beyond 8. For details of the proof see \ref{app:A}.

Theorems 1 and 2 strongly indicate that any $(M+1)$-party generalised GHZ state with $\alpha \neq \frac{1}{2}$ shared between $M$ senders and a single receiver is not good for DDC protocol beyond the classical limit.

A 4-qubit generalized Dicke state \cite{dicke} is representable as
\(\Ket{D_4^r} = \sum\limits_{\mathcal{P}} \alpha_{\mathcal{P}} \mathcal{P}(
\Ket{0}^{\otimes (4-r)} \otimes \Ket{1}^{\otimes r})\),
\noindent where $\mathcal{P}$ denotes the possible permutations of the state with $r$-qubits in the excited state, $\Ket{1}$ and $(4-r)$ in the ground state, $\Ket{0}$. The use of $r=1$ corresponds to the well known four-qubit gW state \cite{Wclass}.
We generate $5 \times 10^3$ gW and $\Ket{D_4^2} $ states Haar uniformly and find that about $15.2 \%$ and $40.5 \%$ states have $N_{max} >8 $ for gW and $\Ket{D_4^2} $ respectively, implying quantum advantage in DDC scheme.

\section{Description of the protocol for DDC network}
The senders, $S_1, S_2, ..., S_M$,  and the receiver, $R$ a priori know what state is shared between them. So, they know $N_{max}$ and the corresponding unitary operators, $\{U^{S_1}_i \otimes U^{S_2}_i\}$ with $i = 1,2,..., N_{max}$, that the senders have to perform to generate orthogonal states at $R$'s end. The receiver, $R$, has the chart of all the encoded orthogonal states corresponding to the different $i$, so that she/he can decode the message by distinguishing the orthogonal states by performing global measurements. Note that in DDC protocols, the members of the set of encoding local unitaries, i.e., $\{U^{S_1}_i\}$ and $\{U^{S_2}_i\}$,  are not independent. This means that one particular sender can not encode her/his message without the knowledge of the unitary used by the other senders. Hence, a classical communication channel is needed between the senders which they use to fix the encoding unitaries depending on the message. 

In the scenario of two senders ($S_1$ and $S_2$) and a single receiver ($R$), the allowed values of $N_{max}$ are $4, 5, 6$ and $8$. If $N_{max}=6$, say, then $S_1$ can send one-bit of information, while $S_2$ can send one-trit of information to $R$, and from the preexisting encoding chart the receiver can separate-out the information coming from different senders. Again from the encoding chart, the receiver can decode the messages coming from different senders.  As another, if $N_{max} = 5$, we can assume that $S_1$ and $S_2$ wish to send one of the options $\{00, 01, 10, 11, 20\}$ away.

In any such case, since a full basis of local unitaries is not used, the encoding is correlated between the senders, i.e, the senders need to know about the other sender's encoding. Note that we are assuming here that classical communication between the senders is not costly, while that between the senders and the receiver remains so. Note also that since the full basis of unitaries is used in the usual, non-deterministic, dense coding, this condition of cost of classical communication between the senders is redundant.

\section{Discussion}
When apriori a non-maximally entangled pure state is shared, the deterministic dense coding has a unique advantage over the prababilistic dense coding schemes from the experimental point of view, since the protocol works  deterministically at the single-copy level, and one does not require to consider many copies of the shared entangled state.
Moreover, quantum communication protocols between two parties have a limited use in
realistic situations.  
A multipartite protocol has  several applications  in day-to-day  scenarios like
reporters sending news to a newspaper editor,  meteorological office getting
information about weather from different measuring sites etc.
We considered transfer of classical information via a quantum channel between multiple senders and a single receiver with certainty, i.e., the deterministic quantum dense coding network.
While in the bipartite scenario, DDC is not possible in qubit systems, we have shown that in the multiparty case, we can have quantum advantage of DDC even using qubit systems, where senders encode their messages by performing local two-dimensional unitary operators.
 We found that in the tripartite case, the entire family of generalized GHZ
states except the GHZ  are not capable for deterministic dense coding,
 while quantum advantage can be obtained when the shared state is the
generalized W state in certain parameter ranges. Such  complementary
results between the generalized GHZ  and  generalized Dicke states
hold also for multiple senders.  This scheme is one of the few examples
where the generalized W state turns out to be advantageous in quantum
information processing tasks over the generalized GHZ state. However, a higher percentage  of states from the GHZ-class are
deterministic dense codeable compared to  that of states from the W
class, in the case of two senders.
As in a bipartite scenario, we find multipartite quantum characteristic like genuine multiparty entanglement, monogamy score of quantum correlation measures, and multiparty
capacity of dense coding in the non-deterministic scenario are unrelated to the capacity of multinode deterministic dense coding.

\section*{Acknowledgments}
The authors acknowledge computations performed at the cluster computing facility of the Harish-Chandra Research Institute, Allahabad, India. This research was supported in part by the `INFOSYS
scholarship for senior students'. 

\appendix
%
%
%
\section{Proof of Theorem 2}

Here we give the proof of Theorem 2. We have the four-qubit gGHZ state, given by $|gGHZ_{S_1 S_2 S_3 R}\rangle = \sqrt{\alpha} |00 00\rangle + \sqrt{1- \alpha}~e^{i \mu}|111 1\rangle$, where $\alpha \in [0,1]$ and $\mu \in [0,2\pi)$. We further assume $\alpha \neq \frac{1}{2}$. 
The most general local encoding can be done by the set of unitaries 
$\{U^{S_1}_i \otimes U^{S_2}_i \otimes U^{S_3}_i \otimes \mathbb{I}_2^R \}$, where $\mathbb{I}_2^R$ is $2\times2$ identity matrix and $U_i^{S_k}$'s are given by Eq. (4) in the main text.

As mentioned before, without any loss of generality, we assume that the first unitary of the  set of orthogonal unitaries is to be identity. Then we look for other unitaries which are orthogonal to identity and among themselves. Given a general unitary $U^{S_1}\otimes U^{S_2} \otimes U^{S_3} \otimes \mathbb{I}_2^R$, orthogonality condition demands
\begin{eqnarray}\label{eq_ortho1}
&&\Bra{gGHZ}  U^{S_1}\otimes U^{S_2} \otimes U^{S_3} \otimes \mathbb{I}_2^R \Ket{gGHZ} = 0.
\end{eqnarray}
For $\alpha \neq \frac{1}{2}$, it reduces to

\begin{equation}
\cos\theta^{S_1}\cos\theta^{S_2}\cos\theta^{S_3} = 0. 
\label{eq_ortho3}
\end{equation}
The solution of the above equation gives rise to the following seven classes of unitaries which can be furthur grouped into three groups.
\begin{center}
\begin{tabular}{ |c|c|c|c|c| }
\hline
Group & Class & $\theta^{S_1}$ & $\theta^{S_2}$ & $\theta^{S_3}$ \\ 
\hline
$ $ & $\mathcal{C}_1$ & arbitrary & arbitrary & $\pi /2$  \\ 

$\mathcal{G}_1 $ & $\mathcal{C}_2$ & arbitrary & $\pi /2$ & arbitrary  \\

$ $ & $\mathcal{C}_3$ & $\pi /2$ & arbitrary & arbitrary  \\
\hline
$ $ & $\mathcal{C}_4$ & arbitrary & $\pi /2$ & $\pi /2$  \\

$\mathcal{G}_2 $ & $\mathcal{C}_5$ & $\pi /2$ & arbitrary & $\pi /2$  \\

$ $ & $\mathcal{C}_6$ & $\pi /2$ & $\pi /2$ & arbitrary   \\
\hline
$\mathcal{G}_3 $ & $\mathcal{C}_7$ & $\pi /2$ & $\pi /2$ & $\pi /2$  \\  

 \hline
\end{tabular}
\end{center}

We observe that the classes in group $\mathcal{G}_i$'s are equivalent in a sense that all classes in the same group has same number of angle variables which are equal to ${\pi}/{2}$, while other nontrivial angle variables are arbitrary, so the orthogonality relations they have to satisfy have the same form. We show that, however differently, we choose unitaries from these three classes, we cannot get more than seven orthogonal unitaries. These seven unitaries along with identity makes eight orthogonal unitaries, which is the classical limit in this case. So there is no quantum advantage in DDC for gGHZ states with $\alpha \neq \frac{1}{2}$.
Now we list some important orthogonality conditions. Note that in many places apart from constraing the angle variables $\{\theta^{S_i}\}$, the phase variables must also be readjusted to obtain the orthogonality. We have already seen in the three qubit case that adjusting the phase variables was trivial. This too holds for the four qubit case. The angle variables are the ones which really control the orthogonality relations. Hence all the orthogonality conditions would be given in terms of the angle variables only, with appropriate readjusting of phase variables $\{x^{{S}_i},y^{{S}_j}\}$  assumed wherever necessary. Before proving that $N^{gGHZ}_{max} \leq 8$, we first derive the orthogonality conditions by choosing unitaries either from different groups or from different classes of the same group. 

\noindent{\bf C1.} Lets us first choose unitaries from different groups.


{\bf C1a.} Orthogonality of a unitary from the group $\mathcal{G}_1$  with another unitary from the group $\mathcal{G}_3$:

Consider a unitary belonging to any class of group $\mathcal{G}_1$. Among its three angle variables coming from three senders, one is ${\pi}/{2}$ and let the others be $\theta^{S_i}$ and $\theta^{S_j}$, with $i \neq j$. Its orthogonality with a unitary from $\mathcal{G}_3$ demands
\begin{equation} \label{eq_g13}
\sin\theta^{S_i}\sin\theta^{S_j} = 0.
\end{equation}

{\bf C1b.} Orthogonality condition between unitaries from groups $\mathcal{G}_2$ and $\mathcal{G}_3$.

The orthogonality of a unitary from $\mathcal{G}_2$ with a unitary from $\mathcal{G}_3$ requires
\begin{equation}\label{eq_g23}
\sin\theta^{S_i} = 0,
\end{equation}
where $\theta^{S_i}$ is the angle variable of the unitary, chosen from any class of $\mathcal{G}_2$. Note that the other angle variables corresponding to other senders in this case are $\pi/2$.

{\bf C1c.} Orthogonality condition between unitaries from $\mathcal{G}_1$ and $\mathcal{G}_2$.


As chosen in {\bf C1a} and {\bf C1b},  unitary $U^1 \text{ and } U^2$ from $\mathcal{G}_1 \text{ and } \mathcal{G}_2$ can respectively be represented by $(\theta^{S_i}, \theta^{S_j}, i\neq j) \text{ and } \bar{\theta}^{S_k}$. Now $k$ can either be equal to $i$ or $j$ or $i \neq j \neq k$. If $k$ is equal either to $i$ or $j$, the orthogonality condition between $U^1$ and $U^2$ reads (assuming $k$ = $i$)

\begin{equation}\label{eq_g12a}
\cos(\bar{\theta}^{S_k} - \theta^{S_i})\sin\theta^{S_j} = 0.
\end{equation}
If $k$ is different from both $i$ and $j$, the condition takes the form
\begin{equation}\label{eq_g12b}
\sin\bar{\theta}^{S_k}\sin\theta^{S_i}\sin\theta^{S_j} = 0.
\end{equation}

\noindent {\bf C2.} Interclass orthogonality relations within the same group.

{\bf C2a.} Unitaries from group $\mathcal{G}_1$:


Let $U^{1a} \text{ and } U^{1b}$ be two unitaries belonging to two different classes of group $\mathcal{G}_1$. Let the angle variables of $U^{1a}$ and $U^{1b}$ respectively be $(\theta^{S_i},\theta^{S_j})$ and 
$(\bar{\theta}^{S_k},\bar{\theta}^{S_j})$. By construction, there exists one common sender, which without loss of generality,  we assume to be $S_j$. The orthogonality relation between $U^{1a}$ and $U^{1b}$ reads

\begin{equation}\label{eq_g11}
\cos(\bar{\theta}^{S_j}-\theta^{S_j})\sin\theta^{S_i}\sin\bar{\theta}^{S_k} = 0.
\end{equation}

{\bf C2b.} Condition for unitaries from group $\mathcal{G}_2$:

Consider two different unitaries $U^{2a}$ and $U^{2b}$ from different classes of group $\mathcal{G}_2$ with   arbitrary angle variables $\theta^{S_i}$ and $\bar{\theta}^{S_j}$ respectively, where $i$ and $j$ are by definition different. Their orthogonality condition is given by
\begin{equation}\label{eq_g22}
\sin\theta^{S_i}\sin\bar{\theta}^{S_j} = 0.
\end{equation}

\noindent {\bf C3.} Intraclass orthogonality relations.

{\bf C3a.} Unitaries from group $\mathcal{G}_1$.


Consider now two unitaries $U^{1a}$ and $U^{1b}$ from a same class of $\mathcal{G}_1$.  The orthogonality relation between $U^{1a}$ and $U^{1b}$ turns out to be

\begin{equation}\label{eq_c11}
\cos(\bar{\theta}^{S_i}-\theta^{S_i})\cos(\bar{\theta}^{S_j}-\theta^{S_j}) = 0,
\end{equation}

where $(\theta^{S_i},\theta^{S_j})$ and $(\bar{\theta}^{S_i},\bar{\theta}^{S_j})$ respectively are angle variables of $U^{1a}$ and $U^{1b}$.

{\bf C3b.} Unitaries from Group $\mathcal{G}_2$:

Consider two different unitaries $U^{2a}$ and $U^{2b}$ from the same class of group $\mathcal{G}_2$ with   arbitrary angle variables $\theta^{S_i}$ and $\bar{\theta}^{S_i}$ respectively. Their orthogonality condition reads
\begin{equation}\label{eq_c22}
\cos(\theta^{S_i} -\bar{\theta}^{S_i}) = 0.
\end{equation}

Based on the above conditions, we list the cases which occur while choosing orthogonal unitaries.

\noindent {\bf Case 1.} We cannot make more than 4 unitaries from a particular class of group $\mathcal{G}_1$ orthogonal.

Using Eq. \eqref{eq_c11}, we can directly write down the nontrivial angle variables, which are not equal to $\pi/{2}$, for the four orthogonal unitaries from a particular class of group $\mathcal{G}_1$ as $\lbrace(\theta^{S_i},\theta^{S_j}),(\theta^{S_i},\pi/2-\theta^{S_j}),(\pi/2-\theta^{S_i},\theta^{S_j}),(\pi/2-\theta^{S_i},\pi/2-\theta^{S_j})\rbrace$. Again from Eq. \eqref{eq_c11}, it is clear that there cannot be any more orthogonal unitaries from that class.

\noindent {\bf Case 2.} We cannot make more than 2 unitaries from a particular class of group $\mathcal{G}_2$ orthogonal.

From Eq. \eqref{eq_c22}, the nontrivial angle variables for the two orthogonal unitaries from a particular class of group $\mathcal{G}_2$ can be obtained. They are $\{(\theta^{S_i}),(\pi/2-\theta^{S_i})\rbrace$. From Eq. \eqref{eq_c22}, it is again clear that there cannot be any more orthogonal unitaries from that class.

\noindent {\bf Case 3.} Seven orthogonal unitaries from group $\mathcal{G}_1$ can be chosen and there is no possibility of choosing even a single one from other groups.

Suppose we construct 4 orthogonal unitaries from any given class of group $\mathcal{G}_1$, say from $\mathcal{C}_1$, in the same way as listed earlier in case {\bf 1}. This leaves us with two other classes, $\mathcal{C}_2 \text{ and } \mathcal{C}_3$ of group $\mathcal{G}_1$. Let us now choose, say, class $\mathcal{C}_2$ from $\mathcal{G}_1$. Eqs. \eqref{eq_g11} and \eqref{eq_c11} suggest we can chose two more orthogonal unitaries from this class with angle variables as $\{ (v_1,0), (v_1-\pi/2, 0) \}$, where $v_1$ can be arbitrary. Naturally, the remaining angle variable is $\pi/2$. Finally, from the remaining class, we can choose one more orthogonal unitary having both the nontrivial angle variables to be $0$. From Eqs. \eqref{eq_g13}, \eqref{eq_g12a} and \eqref{eq_g12b}, we see that any unitary from group $\mathcal{G}_2$ or group $\mathcal{G}_3$ cannot be simultaneously made orthogonal to these seven chosen unitaries of $\mathcal{G}_1$. A typical example of angle variables of seven orthogonal unitaries (with appropriately chosen phases) is given below.
\begin{center}
\begin{tabular}{ |c|c|c|c|c|c| }
\hline
 Group & Class & $\theta^{S_1}$ & $\theta^{S_2}$ & $\theta^{S_3}$ \\ 
\hline
 $\mathcal{G}_1$ & $\mathcal{C}_1$ & $u_1$ & $u_2$ & $\pi/2$  \\ 

 $\mathcal{G}_1$ & $\mathcal{C}_1$ & $u_1-\pi/2$ & $u_2$ & $\pi/2$  \\ 

  $\mathcal{G}_1$ & $\mathcal{C}_1$ & $u_1$ & $u_2-\pi/2$ & $\pi/2$  \\ 

  $\mathcal{G}_1$ & $\mathcal{C}_1$ & $u_1-\pi/2$ & $u_2-\pi/2$ & $\pi/2$  \\ 
\hline
  $\mathcal{G}_1$ & $\mathcal{C}_2$ & $v_1$ & $\pi/2$ & $0$  \\ 
 
  $\mathcal{G}_1$ & $\mathcal{C}_2$ & $v_1-\pi /2$ & $\pi/2$ & $0$  \\ 
\hline
 $\mathcal{G}_1$ & $\mathcal{C}_3$ & $\pi/2$ & $0$ & $0$  \\ 
 \hline
\end{tabular}
\end{center}
In the above table, $u_1, u_2$, and $v_1$ are arbitrary angle variables. Similar construction can be made if we first choose four unitary operators either from $\mathcal{C}_2$ or $\mathcal{C}_3$.

\noindent {\bf Case 4.} We can choose six orthogonal unitaries from group $\mathcal{G}_1$, one from group $\mathcal{G}_2$ and none from the other one.

Like the previous case, we again construct 4 orthogonal unitaries from any given class of group $\mathcal{G}_1$ in the same way as listed earlier in case {\bf 1}. Let in this class, the sender $S_i$ applies a unitary with angle variable $\pi/2$. Now consider a unitary from $\mathcal{G}_2$ whose arbitrary angle variable at $S_i$ is $u_i \neq \pi /2$. The other angle variables are then by definition $\pi/2$. From Eq. \eqref{eq_g12b}, the ortogonality of this unitary with the four unitaries from group $\mathcal{G}_1$ demands $u_i$ to be vanishing. We can now choose only two more unitaries from the remaining two classes of group $\mathcal{G}_1$, both of whose nontrivial angle variables must be equal to $0$ as easily seen from Eqs. \eqref{eq_g12a}, \eqref{eq_g12b}
and \eqref{eq_g11}. From the orthogonality conditions described above, we can conclude that no more orthogonal unitaries can be constructed in this situation. Cases 3 and 4 exhaust all the possible scenarios when one chooses four unitaries from a single class of $\mathcal{G}_1$. A typical example of angle variables of the seven orthogonal unitaries (with appropriately chosen phases) is as follows.

\begin{center}
\begin{tabular}{ |c|c|c|c|c| }
\hline
 Group & Class & $\theta^{S_1}$ & $\theta^{S_2}$ & $\theta^{S_3}$ \\ 
\hline
  $\mathcal{G}_1$ & $\mathcal{C}_1$ & $u_1$ & $u_2$ & $\pi/2$  \\ 

  $\mathcal{G}_1$ & $\mathcal{C}_1$ & $u_1-\pi/2$ & $u_2$ & $\pi/2$  \\ 

 $\mathcal{G}_1$ & $\mathcal{C}_1$ &  $u_1$ & $u_2-\pi/2$ & $\pi/2$  \\ 

  $\mathcal{G}_1$ & $\mathcal{C}_1$ & $u_1-\pi/2$ & $u_2-\pi/2$ & $\pi/2$  \\

  $\mathcal{G}_1$ & $\mathcal{C}_2$ & $0$ & $\pi/2$ & $0$  \\ 

 $\mathcal{G}_1$ & $\mathcal{C}_3$ & $\pi/2$ & $0$ & $0$  \\ 
 
 \hline
  $\mathcal{G}_2$ & $\mathcal{C}_6$ & $\pi/2$ & $\pi/2$ & $0$  \\ 
 \hline
\end{tabular}
\end{center}

\noindent {\bf Case 5.} We can choose five orthogonal unitaries from group $\mathcal{G}_1$ and one each from group $\mathcal{G}_2$ and $\mathcal{G}_3$ and no more. 

Let us now choose five orthogonal unitaries from $\mathcal{G}_1$, where at most two unitaries are taken from a same class.
From Eqs. \eqref{eq_g23} - \eqref{eq_g11}, we can construct only a single unitary $\mathcal{G}_2$ and $\mathcal{G}_3$ which are orthogonal to this set of five unitaries. A typical example of angle variables of the seven orthogonal unitaries is listed in the table below.

 
 .
\begin{center}
\begin{tabular}{ |c|c|c|c|c| }
\hline
 Group & Class & $\theta^{S_1}$ & $\theta^{S_2}$ & $\theta^{S_3}$ \\ 
\hline
  $\mathcal{G}_1$ & $\mathcal{C}_1$ & $u_1$ & $0$ & $\pi/2$  \\ 

  $\mathcal{G}_1$ & $\mathcal{C}_1$ & $u_1-\pi/2$ & $0$ & $\pi/2$  \\ 
\hline
  $\mathcal{G}_1$ & $\mathcal{C}_2$ & $0$ & $\pi/2$ & $u_3$  \\ 

  $\mathcal{G}_1$ & $\mathcal{C}_2$ & $0$ & $\pi/2$ & $u_3-\pi/2$  \\ 
\hline
  $\mathcal{G}_1$ & $\mathcal{C}_3$ & $\pi/2$ & $0$ & $0$  \\ 
 \hline
  $\mathcal{G}_2$ & $\mathcal{C}_4$ & $0$ & $\pi/2$ & $\pi/2$  \\ 
\hline
 $\mathcal{G}_3$ & $\mathcal{C}_7$ & $\pi/2$ & $\pi/2$ & $\pi/2$  \\ 
 \hline
\end{tabular}
\end{center}

\noindent {\bf Case 6.} We can choose four orthogonal unitaries from group $\mathcal{G}_1$, two from $\mathcal{G}_2$, and one from  $\mathcal{G}_3$.

 Let us construct four orthogonal unitaries from $\mathcal{G}_1$. It is possible if we choose two unitaries from a particular class of $\mathcal{G}_1$ and one unitary each, from the remaining two classes. A typical example is as follows.

\begin{center}
\begin{tabular}{ |c|c|c|c|c| }
\hline
 Group & Class & $\theta^{S_1}$ & $\theta^{S_2}$ & $\theta^{S_3}$ \\ 
\hline
  $\mathcal{G}_1$ & $\mathcal{C}_1$ & $u_1$ & $u_2$ & $\pi/2$  \\ 

  $\mathcal{G}_1$ & $\mathcal{C}_1$ & $u_1-\pi/2$ & $u_2$ & $\pi/2$  \\ 
\hline
 $\mathcal{G}_1$ & $\mathcal{C}_2$ & $0$ & $\pi/2$ & $0$  \\ 
\hline
 $\mathcal{G}_1$ & $\mathcal{C}_3$ & $\pi/2$ & $0$ & $0$  \\ 

 \hline
\end{tabular}
\end{center}

Let us try to make these four unitaries orthogonal to unitaries from $\mathcal{G}_2$ and $\mathcal{G}_3$. Using Eqs. \eqref{eq_g12a} and \eqref{eq_g12b}, we can easily see that only three more orthogonal unitaries, two from group $\mathcal{G}_2$ and one from $\mathcal{G}_3$ are possible by setting nontrivial angle variables, along with $u_2$ (given in the above table), to be $0$. A typical configuration for such a case is given below.
\begin{center}
\begin{tabular}{ |c|c|c|c|c| }
\hline
 Group & Class & $\theta^{S_1}$ & $\theta^{S_2}$ & $\theta^{S_3}$ \\ 
\hline
  $\mathcal{G}_1$ & $\mathcal{C}_1$ & $u_1$ & $0$ & $\pi/2$  \\ 

  $\mathcal{G}_1$ & $\mathcal{C}_1$ & $u_1 -\pi/2$ & $0$ & $\pi/2$  \\ 
 
  $\mathcal{G}_1$ & $\mathcal{C}_2$ & $0$ & $\pi/2$ & $0$  \\ 

  $\mathcal{G}_1$ & $\mathcal{C}_3$ & $\pi/2$ & $0$ & $0$  \\ 
\hline
  $\mathcal{G}_2$ & $\mathcal{C}_4$ & $0$ & $\pi/2$ & $\pi/2$  \\ 
 
  $\mathcal{G}_2$ & $\mathcal{C}_6$ & $\pi/2$ & $\pi/2$ & $0$  \\ 
\hline
 $\mathcal{G}_3$ & $\mathcal{C}_7$ & $\pi/2$ & $\pi/2$ & $\pi/2$  \\ 
 \hline
\end{tabular}
\end{center}

\noindent {\bf Case 7.} We can choose four orthogonal unitaries from $\mathcal{G}_2$, three from $\mathcal{G}_1$, and none from  $\mathcal{G}_3$.

We construct two orthogonal unitaries from one particular class of group $\mathcal{G}_2$ in the same way as listed in case {\bf 2}. From Eq. \eqref{eq_g22}, the only way, we can construct more orthogonal unitaries from group 
$\mathcal{G}_2$ is by taking one unitary each from the remainig two classes of $\mathcal{G}_2$ 
and setting their nontrivial angle variables to be $0$. Eq. \eqref{eq_g22} furthur ensures that this is the maxiamal number of orthogonal unitaries, one can choose from $\mathcal{G}_2$. Eq. \eqref{eq_g23} guarantees that this set of four orthogonal unitaries from  $\mathcal{G}_2$ cannot be made orthogonal to any unitary from $\mathcal{G}_3$. Let us extend this set by considering unitaries from $\mathcal{G}_1$. From Eqs. \eqref{eq_g12a} - \eqref{eq_c11}, we clearly see that three more orthogonal unitaries, each from the three different classes of $\mathcal{G}_1$ can be constructed after setting all their non-trivial angle variables set to be $0$. See table below.

\begin{center}
\begin{tabular}{ |c|c|c|c|c| }
\hline
 Group & Class & $\theta^{S_1}$ & $\theta^{S_2}$ & $\theta^{S_3}$ \\ 
\hline
  $\mathcal{G}_1$ & $\mathcal{C}_1$ & $0$ & $0$ & $\pi/2$  \\ 

  $\mathcal{G}_1$ & $\mathcal{C}_2$ & $0$ & $\pi/2$ & $0$  \\ 

  $\mathcal{G}_1$ & $\mathcal{C}_3$ & $\pi/2$ & $0$ & $0$  \\ 
 \hline
  $\mathcal{G}_2$ & $\mathcal{C}_4$ & $u_1$ & $\pi/2$ & $\pi/2$  \\ 

 $\mathcal{G}_2$ & $\mathcal{C}_4$ & $u_1-\pi/2$ & $\pi/2$ & $\pi/2$  \\ 
\hline
  $\mathcal{G}_2$ & $\mathcal{C}_5$ & $\pi/2$ & $0$ & $\pi/2$  \\ 
 
  $\mathcal{G}_2$ & $\mathcal{C}_6$ & $\pi/2$ & $\pi/2$ & $0$  \\ 
 \hline
\end{tabular}

\end{center}

\noindent {\bf Case 8.} We now show that we can at most construct one unitary from each class, and make them orthogonal.

From the orthogonality conditions given in Eqs. \eqref{eq_g13} - \eqref{eq_c22}, we see that there exists an unique configuration of angle variables by which we can obtain seven orthogonal unitaries, one from each class but not more than that, as depicted in the following table.
\begin{center}
\begin{tabular}{ |c|c|c|c|c| }
\hline
 Group & Class & $\theta^{S_1}$ & $\theta^{S_2}$ & $\theta^{S_3}$ \\ 
\hline
  $\mathcal{G}_1$ & $\mathcal{C}_1$ & $0$ & $0$ & $\pi/2$  \\ 

  $\mathcal{G}_1$ & $\mathcal{C}_2$ & $0$ & $\pi/2$ & $0$  \\ 

  $\mathcal{G}_1$ & $\mathcal{C}_3$ & $\pi/2$ & $0$ & $0$  \\ 
\hline

  $\mathcal{G}_2$ & $\mathcal{C}_4$ & $0$ & $\pi/2$ & $\pi/2$  \\ 
  
  $\mathcal{G}_2$ & $\mathcal{C}_5$ & $\pi/2$ & $0$ & $\pi/2$  \\ 
 
 $\mathcal{G}_2$ & $\mathcal{C}_6$ &  $\pi/2$ & $\pi/2$ & $0$  \\ 
\hline

 $\mathcal{G}_3$ & $\mathcal{C}_7$ & $\pi/2$ & $\pi/2$ & $\pi/2$  \\

 \hline
\end{tabular}
\end{center}

Hence the above cases exhaust all possible combinations that one can use to choose orthogonal unitaries for the gGHZ states, with $\alpha \neq \frac{1}{2}$; implying at most eight orthogonal unitaries, including identity. Hence the proof.  \hfill $\blacksquare$

\label{app:A}


\begin{thebibliography}{100}
\bibitem{GHZ_class_state}A state from the GHZ-class can be expressed as
\begin{equation*}
\Ket{\psi_{GHZ}^c} = \sqrt{K}(c_\delta|0\rangle|0\rangle|0\rangle+s_\delta e^{i\varphi}|\varphi_A\rangle|\varphi_B\rangle|\varphi_C\rangle),
\end{equation*}
where $ \Ket{\varphi_A}=c_{\alpha}\Ket{0}+s_{\alpha}\Ket{1}$,
$\Ket{\varphi_B}=c_{\beta}\Ket{0}+s_{\beta}\Ket{1}$,
$\Ket{\varphi_C}=c_{\gamma}\Ket{0}+s_{\gamma}\Ket{1}$,
 $K=(1+2 c_\delta s_\delta c_\alpha c_\beta c_\gamma c_\varphi)^{-1} \in
(1/2,\infty)$ is a normalization factor, and $\delta \in (0,\pi/4]$, $\alpha,\beta,\gamma \in (0,\pi/2]$ and $\varphi \in
[0,2\pi)$. Here $c_{x} = \cos x$, $s_{x} = \sin x$, $x = \alpha,\beta,\gamma, \phi, \delta$.
\bibitem{W_class_state} A member state of the W-class, up to a local 
unitary transformation, can be written as
\begin{equation*}
\Ket{\psi_W} = \sqrt{a} \Ket{001} +\sqrt{b} \Ket{010} + \sqrt{c} \Ket{100} +\sqrt{d} \Ket{000},
\end{equation*}
where $a, b, c >0$, and $d = 1- (a + b+ c) \geq 0$.
\bibitem{ent_multi} A pure multipartite state is genuinely multiparty entangled if it is not separable across any bipartition. The generalized geometric measure \cite{GGM, GM} quantifies the content of genuine multiparty entanglement and is defined by
\begin{equation*}
g(\ket{\psi}) = 1 - \mbox{max} |\langle \phi | \psi \rangle |^2,
\end{equation*}
where the maximization is performed among the set of states, $\{\ket{\phi}\}$, which are not genuinely multipartite entangled.
\bibitem{neg_mono} The squared negativity monogamy score for a tripartite state $\ket{\psi_{ABC}}$ is defined as \cite{monogamyreview}
\begin{equation*}
\delta_{\mathcal{N}^2} = \mathcal{N}^2(\ket{\psi_{A:BC}}) - \mathcal{N}^2(\rho_{AB}) -\mathcal{N}^2(\rho_{AC}),
\end{equation*}
where \(\ket{\psi_{A:BC}}\) denotes the \(A:BC\) bipartition of \(\ket{\psi_{ABC}}\) and \(\rho_{AB}\) (\(\rho_{AC}\)) is the reduced state \(AB\) (\(AC\)) of  \(\ket{\psi_{ABC}}\). Negativity \cite{neg_papers}, $\mathcal{N}(\rho_{AB})$, of a bipartite state, $\rho_{AB}$, is defined as the absolute sum of negative eigenvalues of the partial transposed state $\rho_{AB}^{T_A}$ \cite{PPT} of $\rho_{AB}$.
\bibitem{capa_def} Multiparty dense coding capacity of a $(M+1)$-party shared state, $\rho_{1 2 \cdots M+1}$, between $M$ senders and a single receiver is given by \cite{capacity_multi}
\begin{equation*}
C(\rho_{1 \cdots M M+1}) = \log_2 d_1 \cdots d_M + S(\rho_{M+1}) - S(\rho_{1 2 \cdots M+1}),
\end{equation*}
where $d_1, \cdots, d_M$ are the dimensions of senders' subsystems and $\rho_{M+1}$ is the  local density matrix of the single receiver.
\end{thebibliography}
\end{document}